\documentclass[11pt,preprintnumbers,aps,amssymb,nofootinbib,amsmath,superscriptaddress]
{revtex4}
\usepackage{epsfig,epsf}
\usepackage{bm} 
\newcommand{\beq}{\begin{equation}}
\newcommand{\beql}[1]{\begin{equation}\label{#1}}
\newcommand{\eeq}{\end{equation}}
\def\bal#1\gal{\begin{align}#1\end{align}}
\newcommand{\ball}[1]{\bal\label{#1}}
\newcommand{\eq}[1]{(\ref{#1})}
\newcommand{\fig}[1]{Fig.~\ref{#1}}
\renewcommand{\sec}[1]{Sec.~\ref{#1}}
\newcounter{topiccounter}
\renewcommand{\b}[1]{{\bm #1}} 
\newcommand{\unit}[1]{\hat {{\bm #1}}} 

\newcommand{\e}{\varepsilon}
\newcommand{\aver}[1]{\left\langle #1 \right\rangle}

\begin{document}


\title{On the role of magnetic field in photon excess in heavy ion collisions}

\author{Kirill Tuchin}

\affiliation{Department of Physics and Astronomy, Iowa State University, Ames, IA 50011}

\date{\today}


\begin{abstract}

Synchrotron photon spectrum in heavy-ion collisions is computed taking into account the spatial and temporal structure of magnetic field. It is found that a significant fraction of photon excess in heavy-ion collisions in the region $k_\bot=1-3$~GeV can be attributed to the synchrotron radiation. Azimuthal anisotropy of the synchrotron photon spectrum is characterized by the ``flow" coefficients $v_2=4/7$ and $v_4=1/10$ that are independent of photon momentum and centrality. 

\end{abstract}

\maketitle


\section{Introduction}\label{sec:intr}

One of the outstanding puzzles in the phenomenology of the heavy-ion collisions is excess of photons at low transverse momenta above the photon spectrum in $pp$ collisions scaled in proportion to the number of binary nucleon collisions \cite{Adare:2014fwh}. Another related problem  is  large azimuthal asymmetry of the photon spectrum \cite{Adare:2011zr}.  The traditional phenomenological approaches   \cite{Linnyk:2013wma,vanHees:2011vb,Shen:2013vja} has recently  improved their agreement with the data, although the discrepancy is not completely eliminated. A novel mechanism of photon production was proposed in \cite{Basar:2012bp}.  In \cite{Tuchin:2012mf,Tuchin:2010gx} synchrotron photon radiation by the quark-gluon plasma was investigated and found to give an important contribution to the total photon spectrum.  In this paper I go beyond the constant field approximation, employed in \cite{Tuchin:2012mf,Tuchin:2010gx}, and compute the synchrotron photon spectrum taking into account the realistic space-time structure of the electromagnetic field.

Electromagnetic field is initially generated by the valance charges of the colliding ions, but at very early times gives way to the induced field generated by the electric currents in the produced matter and travels along with the expanding system \cite{Tuchin:2013ie,Tuchin:2013apa}. The proof of its existence relies only upon the applicability of the effective hydrodynamic description of the final state. Important features of this field are: (i) Its strength at time $t$ is determined only by the collision impact parameter $b$ and the electrical conductivity $\sigma$. It  does not explicitly depend on the collision energy. Rather, energy dependence comes through the variation of $\sigma$ with the temperature $T$. (ii) Its dominant component is magnetic field perpendicular to the event plane \cite{Kharzeev:2007jp}.  

Motion of charged particles of energy $\e$ and charge $e$  in magnetic field $B$ is quantized, with the distance between the nearby Landau levels being on the order of $\omega_B=eB/\e$. However, if $eB\ll  \e^2$, the quantization effect is small. In a thermal medium of temperature $T$ this condition becomes $eB\ll T^2$.  The peak strength of magnetic field at the collision energy $\sqrt{s_{NN}}=200$~GeV is estimated to be $eB= m_\pi^2$ implying that one can treat the synchrotron emission in the quasi-classical approximation. This argument is supported by an explicit calculation in \cite{Tuchin:2012mf}, where I showed that the number of Landau levels contributing to the synchrotron radiation at the field strength $eB= m_\pi^2$ is on the order of a hundred. 

It is well-known, that the synchrotron radiation is emitted over a short time $\Delta t\sim \omega_B^{-1}(m/\e)^3$ \cite{Landau:1982dva}, which is much shorter than the characteristic time of  the magnetic field variation $t_B\sim |B/\dot B|$. This allows me to treat the synchrotron radiation as an adiabatic process, viz.\ to substitute the expression for the time-dependent field \eq{f49} into the emission rate in a constant $B$ \eq{b73}, which is well-known in the literature.  

The results of my calculation indicate that although the synchrotron radiation cannot be responsible for all the observed photon excess, it gives a  significant contribution at photon energies $k_\bot=1-3$~GeV in the central rapidity region.  Since radiation in the direction of the magnetic field vanishes, the synchrotron spectrum exhibits strong azimuthal asymmetry with the following Fourier coefficients: $v_2=4/7$, $v_4=1/10$. This may explain the strong elliptic flow of prompt photons observed in the data \cite{Adare:2011zr}. 

The paper is structured as follows: In \sec{sec:b} an analytic expression for the synchrotron spectrum emitted by a relativistic charge is presented. In \sec{sec:f} I compute the photon spectrum radiated by the quark-gluon plasma during its entire life-time using the explicit space-time dependence of magnetic field discussed in Appendix. The results are shown in \fig{fig1}, \fig{fig2} and \fig{fig3}. In \sec{sec:concl} the summary is presented. 

\section{Photon radiation by a relativistic quark}\label{sec:b}

Consider a relativistic quark or antiquark of energy $\e_0$, velocity $\b v_0$ and electric charge $q_fe$ moving in a plane perpendicular to magnetic field $\b B_0$. I will call the corresponding reference  frame $K_0$.  Emission rate of photon of energy $\omega_0$ and momentum $\b k_0 = \omega_0 \b n_0$  is given by \cite{Berestetsky:1982aq}
\bal
d \dot w_0=& \frac{\alpha q_f^2}{(2\pi)^2}\frac{d^3k_0}{\omega_0}\int_{-\infty}^{+\infty} d\tau \exp\left\{-\frac{i\e_0}{\e'_0}\omega_0 \tau \left [1-\b n_0\cdot \b v_0+\left(\frac{q_feB_0}{\e_0}\right)^2\frac{\tau^2}{24}\right]\right\}\,\nonumber\\
&\times
\left[ -\frac{\e_0'^2+\e_0^2}{4\e_0'^2}\left(\frac{q_feB_0}{\e_0}\right)^2\tau^2-\frac{m^2}{\e_0\e'_0}\right]\label{b73}\,,
\gal
where $\e_0'= \e_0-\omega_0$. 

Consider now another reference frame $K$ where quarks have an arbitrary direction of momentum. 
Let the $y$-axis be in the magnetic field direction $\b B= B\unit y$ and $\b V = V\unit y$ be the velocity of $K$ with respect to $K_0$. Then the Lorentz transformation reads
\bal
&p_{x0}= p_x\,, \quad 0=p_{y0}=  \gamma(p_y+V\e)\,,\quad p_{z0}= p_z\,,\quad \e_0= \gamma(\e+Vp_y)\,.\label{d11}\\
& k_{x0}= k_x\,,\quad k_{y0}= \gamma(k_y+V\omega)\,,\quad k_{z0}=k_z\,,\quad \omega_0= \gamma(\omega+Vk_y)\,.\label{d12}\\
&\b B_0= \b B\,, \label{d13}
\gal
where $\gamma= 1/\sqrt{1-V^2}$. It follows from the second equation in \eq{d11} that
\ball{d15}
V= -\frac{p_y}{\e}
\gal 
and 
\ball{d17}
\e_0= \sqrt{\e^2-p_y^2}\,,\qquad \omega_0= \frac{\omega \e- p_y k_y}{\sqrt{\e^2-p_y^2}}\,.
\gal
  Using the boost invariance of $k\cdot p$ we get 
\ball{d19}
1-\b n_0\cdot \b v_0 = \frac{\omega \e}{\omega_0\e_0}(1-\b n\cdot \b v)\,, 
\gal
accurate up to the terms of the order $m^2/\e^2$. Transformation of the photon emission rate reads \cite{Tuchin:2013bda}
\ball{d21}
\frac{d\dot w}{d\Omega d\omega}= \frac{1}{\gamma^2(1+V\cos\theta)}\frac{d\dot w_0}{d\Omega_0 d\omega_0}=\frac{\omega \e_0}{\e \omega_0}\frac{d\dot w_0}{d\Omega_0 d\omega_0} \,,
\gal
where  $\theta$  is the angle between the photon momentum $\b k$ and the magnetic field, i.e.\ $\cos\theta = n_y$, and $\Omega$ is the corresponding solid angle. In the last step I used \eq{d15} and \eq{d17}. $d\dot w_0$ in the right-hand-side of \eq{d21} is given by \eq{b73}.

\section{Electromagnetic radiation by plasma}\label{sec:f}

\subsection{Photon rate per unit volume}

Quark-gluon plasma in magnetic field radiates photons   into a solid angle $d\Omega$ in the frequency interval  ($\omega$, $\omega+d\omega$) with  the following rate 
\ball{f11}
\frac{dN}{dtd\Omega d\omega}= 2N_c\sum_{f}\int \frac{d\mathcal{V} d^3p}{(2\pi)^3} f(\e)[1-f(\e')]
\frac{d\dot w}{d\Omega d\omega}\,,
\gal
where $\mathcal{V}$ stands for the volume, the sum runs over the quark and anti-quark flavors and the quark/antiquark distribution function in plasma at temperature $T$ reads
\ball{f13}
f(\e)= \frac{1}{e^{\e/T}+1}\,.
\gal

Introduce now a Cartesian reference frame span by three unit vectors $ \b e_1, \b e_2, \b n$, such that vector $\b B$ lies in  plane span by $\b e_1$ and $ \b n$. In terms of the polar and azimuthal angles $\chi$ and $\psi$ we can write
\bal
&\b v= v(\cos\chi \, \b n+ \sin\chi \cos\psi \,\b e_1+\sin\chi \sin\psi \,\b e_2)\,,\label{f15}\\
& \b B= B(\cos\theta\, \b n_1+\sin\theta\, \b e_1)\,.\label{f16}
\gal
Then, 
\bal
&p_y= \frac{\b p\cdot \b B}{B}= \e v(\cos\chi\cos\theta+\sin\chi \cos\psi \sin\theta)\,,\label{f19}\\
&k_y= \frac{\b k\cdot \b B}{B}=\omega \cos\theta\,,\label{f20}\\
& \b n\cdot \b v = v\cos\chi\,.\label{f21}
\gal

Quarks moving in plasma parallel to the magnetic field direction do not radiate due to the vanishing Lorentz force. Bearing in mind  that at high energies quarks radiate mostly into a narrow cone with the opening angle $\chi \sim m/\e$, we conclude that photon radiation at angles $\theta\lesssim m/\e$ can be neglected. Thus, expanding at small $\chi$ we obtain from \eq{d17},\eq{f19}
\ball{f23}
\e_0\approx   \e \sin\theta\,,\quad \omega_0\approx \omega \sin\theta\,, \qquad \theta >\frac{m}{\e}\,.
\gal
Omission of terms of order $m/\e$ is  consistent with the accuracy of \eq{b73}.  In view of \eq{f23}, dependence of the integrand of \eq{f11} on angle $\chi$ comes about only in \eq{d19}, viz.\ 
\ball{f25}  
1-\b n_0\cdot \b v_0= \frac{1}{\sin^2\theta}\left(1-\cos\chi +\frac{m^2}{2\e^2}\right)\,,
\gal
while it is $\psi$-independent.

To integrate over the quark/antiquark  momentum directions $do= d\cos\chi\, d\psi$  we write \eq{f11}  as 
\bal
\frac{dN}{dt d\Omega d\omega}= &\frac{2N_c}{(2\pi)^3}\sum_{f}\int d\mathcal{V}\int_\omega^\infty d\e\, \e^2  f(\e)[1-f(\e')]
 \int do\, \frac{d\dot w}{d\Omega d\omega}\,,
\label{f31}
\gal
substitute \eq{d21} and  \eq{b73} and integrate first over $do$ and then over $\tau$ with the following result (see details in \cite{Berestetsky:1982aq}):
\bal
\int do\, \frac{d\dot w_{T}}{d\Omega d\omega}= &-\frac{\alpha q_f^2 m^2}{\e^2}\sin^2\theta\left\{ \int_{z_\theta}^\infty \text{Ai}(z')dz' +(\sin\theta)^{2/3} \left( \frac{\e}{\e'}\right)^{1/3} \left( \frac{\omega_B}{\omega}\right)^{2/3}\frac{\e^2+\e'^2}{m^2} \text{Ai}'(z_\theta)
\right\}\label{f41}\,,
\gal
where $\omega_B= q_feB/\e$ and 
\ball{f44}
z_\theta=\left( \frac{\e}{\e'}\right)^{2/3} \left( \frac{\omega}{\omega_B}\right)^{2/3} \frac{m^2}{\e^2\sin^{8/3}\theta}\,.
\gal

\subsection{Photon spectrum}

Spatial and temporal dependence of the photon production rate \eq{f31} comes about from the corresponding dependence of the background magnetic field. The explicit form of magnetic field is given in Appendix~\ref{appB}. Neglecting small variations of magnetic field strength in the transverse plane, integration over the time and volume of plasma yields the total photon multiplicity spectrum radiated into a unit solid angle 
\bal
\frac{dN}{d\Omega d\omega}= &\frac{2N_c}{(2\pi)^3}S\sum_{f}\int_0^{t_f} dt \int_{-t}^t dz   \int_\omega^\infty d\e\, \e^2  f(\e)[1-f(\e')]
 \int do\, \frac{d\dot w}{d\Omega d\omega}\,,
\label{f55}
\gal
with \eq{f49} substituted into \eq{f41},\eq{f44} and the overlap area $S$ of two spherical nuclei of radius $R_A$ given by
\ball{f53}
S=R_A^2\,[2\arccos(b/2R_A)-\sin(2\arccos(b/2R_A)]\,.
\gal

The experimental observable is the photon multiplicity at a given transverse momentum $k_\bot$, azimuthal angle $\phi$ and rapidity $y$ with respect to the collisions axis. It reads
\ball{f60}
\frac{dN(k_\bot,\phi, y)}{k_\bot dk_\bot d\phi dy}= \frac{dN(\omega,\theta)}{\omega d\omega d\Omega}\,,
\gal
where $\omega= k_\bot \cosh y$ and $\cos\theta= \sin\phi/\cosh y$. It is usually represented as the cosine Fourier series
\ball{f63}
\frac{dN(k_\bot,\phi,y)}{k_\bot dk_\bot d\phi dy}= \aver{\frac{dN}{d^2k_\bot dy}}_\phi\left( 1+ \sum_{n=1}^\infty 2v_n \cos(n\phi) \right)\,,
\gal
where the azimuthally averaged multiplicity is given by
\bal
&\aver{\frac{dN}{d^2k_\bot dy}}_\phi = \frac{1}{2\pi}\int_0^{2\pi} \frac{dN}{d^2k_\bot dy}\, d\phi\,,\label{f66}
\gal
and the ``flow" coefficients by
\bal
&v_n=\frac{1}{2\pi} \int_0^{2\pi}\frac{dN}{d^2k_\bot dy}\, \cos(n\phi) d\phi \aver{\frac{dN}{d^2k_\bot dy}}_\phi^{-1}\,.\label{f67}
\gal

\begin{figure}[ht]
      \includegraphics[height=5cm]{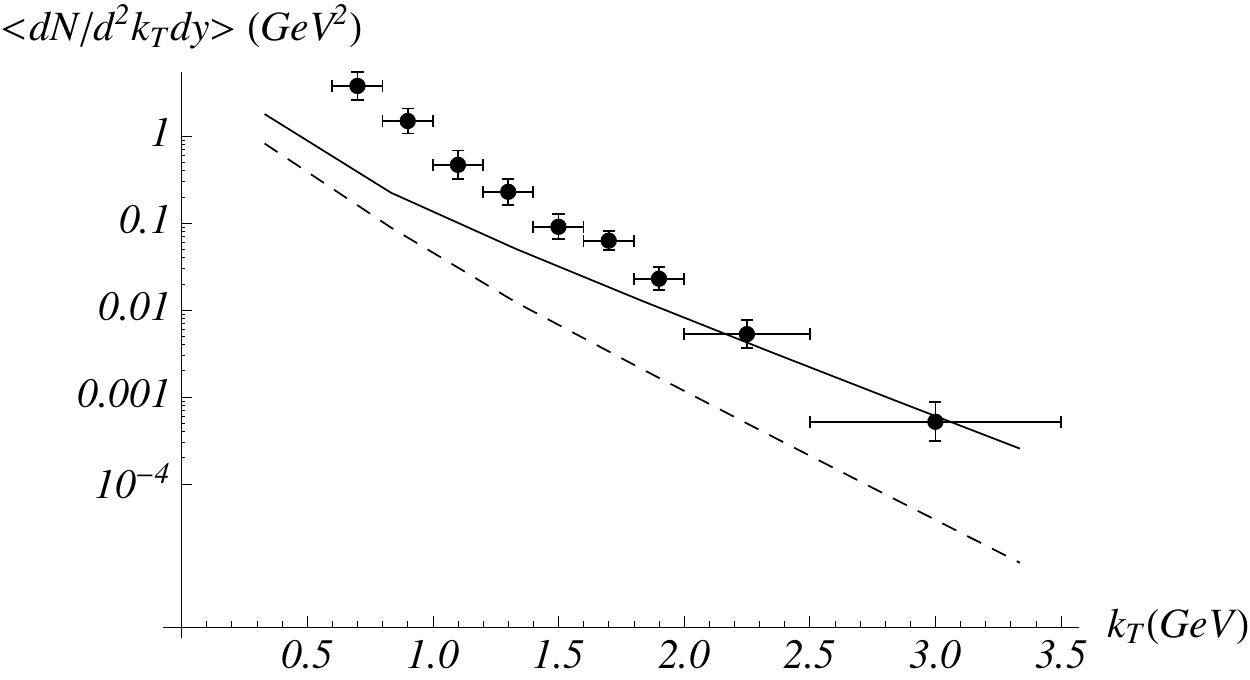} 
  \caption{Spectrum of synchrotron photons averaged over the azimuthal angle versus photon transverse momentum $k_\bot$ at rapidity $y=0$ and centrality $0\%-20$\% ($b=4.3$~fm \cite{Kharzeev:2000ph}). Solid line: $T=400$ MeV, dashed line: $T=200$ MeV. Data is from \cite{Adare:2014fwh}. }
\label{fig1}
\end{figure}

\begin{figure}[ht]
      \includegraphics[height=5cm]{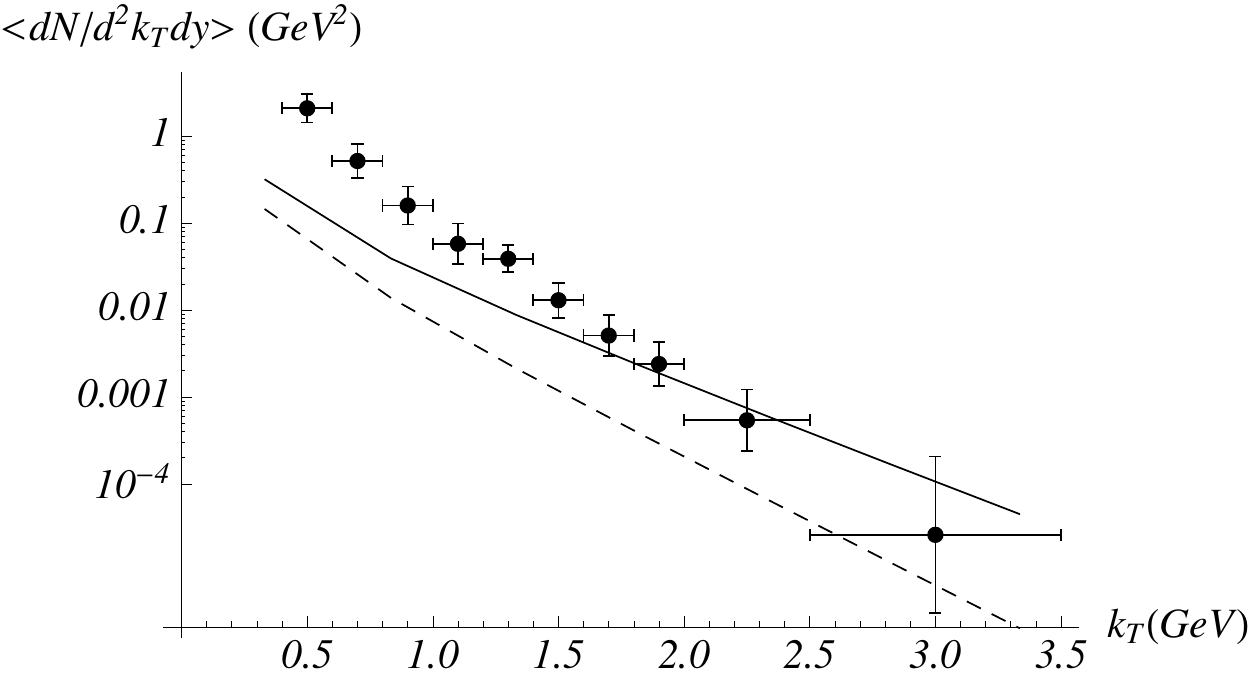} 
  \caption{Spectrum of synchrotron photons averaged over the azimuthal angle versus photon transverse momentum $k_\bot$ at rapidity $y=0$ and centrality $40\%-60$\% ($b=10.2$~fm \cite{Kharzeev:2000ph}). Solid line: $T=400$ MeV, dashed line: $T=200$ MeV. Data is from \cite{Adare:2014fwh}.}
\label{fig2}
\end{figure}

In \fig{fig1} and \fig{fig2} I display the spectrum of synchrotron plasma radiation over time $t\le t_f=10$~fm at different temperatures and centralities. One can see that at low $k_\bot$ synchrotron photons cannot account for the bulk of the photon excess. However, is contributes a substantial fraction of photons at $k_\bot=2-3$~GeV. 


\begin{figure}[ht]
      \includegraphics[height=5cm]{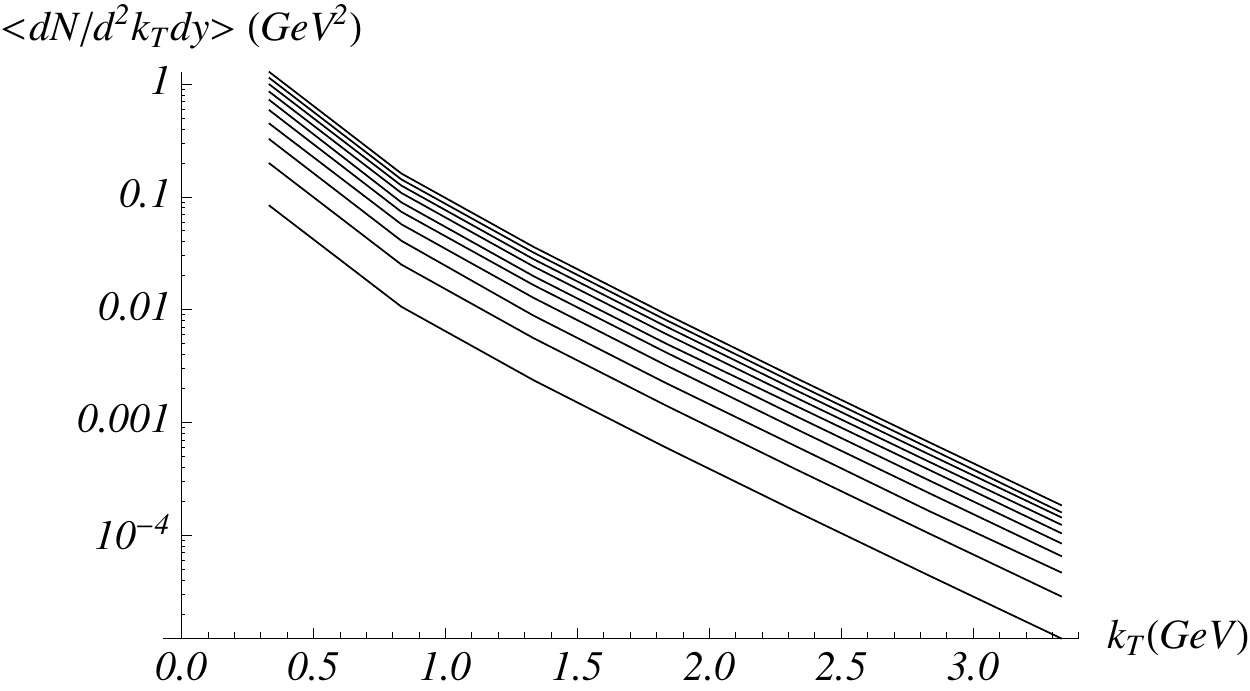} 
  \caption{Time evolution of the photon spectrum (emitted by $u$ and $\bar u$ quarks) from $t=1$~fm (the lowest line) to $t=10$~fm (the highest line) in time increments of 1~fm. $T=400$~MeV, $0\%-20\%$ centrality, $y=0$.}
\label{fig3}
\end{figure}

\fig{fig3} shows the time evolution of the photon spectrum. It is interesting to note that although the spectrum grows fastest at early times it is still increasing even near the freeze-out time $t_f$. This is because the photon spectrum is proportional to $B^{2/3}$ (see \eq{fd11}) while magnetic field decreases as $B\sim 1/t^2$, so that the spectrum is proportional to $1/t_f^{1/3}$. It seems to me that taking into account the time-dependence of plasma temperature and conductivity will lead to a faster decrease of the photon emission rate with time, as can be inferred from \eq{fd11}. 

Concerning the Fourier coefficients \eq{f67}, the ones with  odd indexes vanish $v_{2k+1}=0$, $k=0,1,2,\ldots$, while the ones with even indexes $v_{2k}$ rapidly decrease with increase of $k$. Two largest coefficients are $v_2=0.57$ and $v_4=0.10$. They turned out to be independent of $k_\bot$ and centrality. I will explain this behavior in the next subsection. Here I would like to note, that in view of the results shown in  \fig{fig1} and \fig{fig2}, large elliptic flow of photons observed in \cite{Adare:2011zr} seems to be at least partially due to the strong azimuthal asymmetry of the synchrotron radiation, which is in turn  a consequence of the $\b v\times \b B$ form of the Lorentz force.

\subsection{Photon spectrum at high $k_\bot$}

Analytical expressions for the photon spectrum can be found for photons with $k_\bot\gg T$, which in fact applies to most of the phenomenologically relevant photons.  In this limit we approximate   $f(\e)\approx e^{-\e/T}$ and $z_\theta\ll 1$. Keeping in \eq{f55} only the leading terms in $z_\theta$ and neglecting $m$ compared to $T$ we obtain
\ball{fd11}
\frac{dN}{d^2k dy}= \alpha\frac{2N_c}{(2\pi)^3}\frac{\Gamma(2/3)}{3^{1/3}\Gamma(1/3)}(\sin\theta)^{8/3}e^{-k/T}T^{2/3}\sum_f\int d\mathcal{V}\int_0^{t_f} dt\, (q_f eB)^{2/3} \,.
\gal
Substituting into \eq{f66} we derive for the average photon multiplicity
\ball{fd13}
\aver{\frac{dN}{d^2k_\bot dy}}_\phi =\alpha\frac{2N_c}{(2\pi)^3}  \frac{\Gamma(11/6)}{3\cdot 6^{1/3}\Gamma(7/6)\Gamma(7/3)}e^{-k/T}T^{2/3}\sum_f\int d\mathcal{V}\int_0^{t_f} dt\, (q_feB)^{2/3}\,,
\gal
while the Fourier coefficients follow from \eq{f67}: 
\bal
v_2&= \int_{-\pi/2}^{\pi/2}\cos(2\phi) (\cos\phi)^{8/3}d\phi\Bigg/ \int_{-\pi/2}^{\pi/2} (\cos\phi)^{8/3}d\phi = \frac{4}{7}\,,\label{fd15}\\
v_4&= \int_{-\pi/2}^{\pi/2}\cos(4\phi) (\cos\phi)^{8/3}d\phi\Bigg/ \int_{-\pi/2}^{\pi/2} (\cos\phi)^{8/3}d\phi = \frac{1}{10}\,.\label{fd16}
\gal
Eq.~\eq{fd13} gives a reasonable  approximation for the high $k_\bot$ tail of the photon spectrum. Especially striking is the agreement between \eq{fd15} and \eq{fd16} and the values of $v_2$ and $v_4$  cited in the previous subsection. Apparently, the dominant contribution to the azimuthal angle integration arises at high $k_\bot$. This fact then explains independence of the Fourier coefficients on $k_\bot$, $T$, $B$ and other parameters.

\section{Conclusions}\label{sec:concl}

In this paper I computed the synchrotron photon spectrum in heavy-ion collisions taking into account the spatial and temporal structure of magnetic field. Results obtained in this paper indicate that a significant fraction of photon excess in heavy-ion collisions in the region $k_\bot=1-3$~GeV can be attributed to the synchrotron radiation. Azimuthal anisotropy is characterized by the ``flow" coefficients $v_2=4/7$ and $v_4=1/10$ that are independent of photon momentum and centrality. 

Throughout the paper I assumed that plasma temperature and electrical conductivity  are time-independent which allowed me to use the the analytical expressions for magnetic field \eq{f46}-\eq{f49}. This approach should give rather accurate estimate of the photon spectrum because time variation of temperature and electrical conductivity is rather mild. For example, in the Bjorken scenario $\sigma$, $T\propto t^{-1/3}$  \cite{Bjorken:1982qr}. Nevertheless, a more accurate approach should incorporate a realistic flow of plasma, see e.g.\ \cite{McLerran:2013hla,Zakharov:2014dia}.

\acknowledgments
I would like to thank Sanshiro Mizuno for providing the experimental data. 
This work  was supported in part by the U.S. Department of Energy under Grant No.\ DE-FG02-87ER40371.

\appendix

\section{A model for magnetic field in heavy-ion collisions}\label{appB}

Analytic expression for electromagnetic field created in heavy-ion collisions was found in \cite{Tuchin:2013ie,Tuchin:2013apa}. It is a sum over $Z$ point charges  moving in the positive $z$ direction and $Z$ point charges moving in the opposite direction. Equations simplify in the relativistic limit $\gamma\sigma b\gg 1$. In this case magnetic field created at the origin by a point charge $e$ moving along the positive $z$-axis at transverse distance $b$ reads 
\ball{f46}
\b B= \frac{e}{2\pi}\unit \phi\left( \frac{\gamma b}{2(b^2+\gamma^2 t^2)^{3/2}}+ \frac{b\sigma}{4t^2}e^{-\frac{b^2\sigma}{4t}}\right)\,.
\gal
The first term in the bracket is the boosted Coulomb field in vacuum, while the second term is the field induced in the medium. The quark-gluon system is released from the nuclear wave-functions by  $t\sim 1/Q_s\sim 0.2$~fm, where $Q_s$ is the saturation momentum. By that time the Coulomb term is negligible so that the field in the medium is determined only by $b$ and $\sigma$. Therefore, the total magnetic field  is given by 
\begin{align}\label{f47}
 \b B= \frac{e}{2\pi}\left[ \theta(t-z)\sum_{a=1}^Z\frac{\sigma(\b b/2- \b b_a)}{4(t-z)^2}e^{-\frac{\sigma(\b b/2- \b b_a)^2}{4(t-z)}}+\theta(t+z)\sum_{a=1}^Z\frac{\sigma(\b b/2- \b b_a)}{4(t+z)^2}e^{-\frac{\sigma(\b b/2- \b b_a)^2}{4(t+z)}}\right]\,,
\end{align}
where $\b b_a$'s are the proton transverse coordinates, $\b b $ is the impact parameter,  $z$ is the longitudinal position, $\theta$ is a step-function and  $\alpha=e^2/4\pi$ is the fine structure constant. At large $Z$ magnetic field \eq{f47} is approximately isotropic in the xy-plane (i.e.\ in the plane transverse to the collision axis) 
  and can be well described by the following model
\begin{align}\label{f49}
\b B = \frac{eZ}{2\pi}\unit y \left[\theta(t-z)\frac{\sigma(R_p+ b/2)}{4(t-z)^2} e^{-\frac{(R_p+ b/2)^2\sigma}{4(t-z)}}
+\theta(t+z)\frac{\sigma(R_p+ b/2)}{4(t+z)^2}e^{-\frac{(R_p+ b/2)^2\sigma}{4(t+z)}}\right]\,.
\end{align} 
 Quantum uncertainty of a proton position is accounted for by a finite parameter $R_p=1$~fm \cite{Bzdak:2011yy}. 


\end{document}